\documentclass[pdflatex,sn-mathphys-num]{sn-jnl}

\usepackage{graphicx}%
\usepackage{multirow}%
\usepackage{amsmath,amssymb,amsfonts}%
\usepackage{amsthm}%
\usepackage{mathrsfs}%
\usepackage[title]{appendix}%
\usepackage{xcolor}%
\usepackage{textcomp}%
\usepackage{manyfoot}%
\usepackage{booktabs}%
\usepackage{algorithm}%
\usepackage{algorithmicx}%
\usepackage{algpseudocode}%
\usepackage{listings}%

\begin{document}

\title[Article Title]{A massive and evolved slow-rotating galaxy in the early Universe}

\author*[1]{\fnm{Ben} \sur{Forrest}}\email{bforrest@ucdavis.edu}

\author[2]{\fnm{Adam} \sur{Muzzin}}

\author[3]{\fnm{Danilo} \sur{Marchesini}}

\author[3]{\fnm{Richard} \sur{Pan}}

\author[3]{\fnm{Nehir} \sur{Ozden}}

\author[4]{\fnm{Jacqueline} \sur{Antwi-Danso}}

\author[5]{\fnm{Wenjun} \sur{Chang}}

\author[6]{\fnm{M. C.} \sur{Cooper}}

\author[2]{\fnm{Adit H.} \sur{Edward}}

\author[7]{\fnm{Percy} \sur{Gomez}}

\author[8]{\fnm{Lucas} \sur{Kimmig}}

\author[9,1]{\fnm{Brian C.} \sur{Lemaux}}

\author[10]{\fnm{Ian} \sur{McConachie}}

\author[11]{\fnm{Allison} \sur{Noble}}

\author[8]{\fnm{Rhea-Silvia} \sur{Remus}}

\author[6,12]{\fnm{Stephanie M.} \sur{Urbano Stawinski}}

\author[13]{\fnm{Gillian} \sur{Wilson}}

\author[13]{\fnm{M. E.} \sur{Wisz}}

\affil[1]{\orgdiv{Department of Physics and Astronomy}, \orgname{University of California, Davis}, \orgaddress{\street{One Shields Avenue}, \city{Davis}, \state{CA} \postcode{95616}, \country{USA}}}

\affil[2]{\orgdiv{Department of Physics and Astronomy}, \orgname{York University}, \orgaddress{\street{4700 Keele Street}, \city{Toronto}, \postcode{MJ3 1P3}, \state{ON}, \country{Canada}}}

\affil[3]{\orgdiv{Department of Physics and Astronomy}, \orgname{Tufts University}, \orgaddress{\street{574 Boston Avenue}, \city{Medford}, \state{MA}  \postcode{02155},  \country{USA}}}

\affil[4]{\orgdiv{David A. Dunlap Department of Astronomy \& Astrophysics}, \orgname{University of Toronto}, \orgaddress{\street{50 St George Street}, \city{Toronto}, \postcode{M5S 3H4}, \state{ON}, \country{Canada}}}

\affil[5]{\orgdiv{Department of Physics and Astronomy}, \orgname{University of California, Riverside}, \orgaddress{\street{900 University Avenue}, \city{Riverside}, \state{CA} \postcode{92521},  \country{USA}}}

\affil[6]{\orgdiv{Department of Physics and Astronomy}, \orgname{University of California, Irvine}, \orgaddress{\street{4129 Frederick Reines Hall}, \city{Irvine}, \state{CA} \postcode{92697}, \country{USA}}}

\affil[7]{\orgdiv{} \orgname{W. M. Keck Observatory}, \orgaddress{\street{65-1120 Mamalahoa Hwy.}, \city{Kamuela}, \state{HI} \postcode{96743}, \country{USA}}}

\affil[8]{ Universit\"ats-Sternwarte M\"unchen, Fakult\"at f\"ur Physik,
Ludwig-Maximilians-Universit\"at M\"unchen, Scheinerstr.\ 1, D-81679
M\"unchen, Germany}

\affil[9]{\orgdiv{Gemini Observatory}, \orgname{NSF's NOIRLab}, \orgaddress{\street{670 N. A'ohoku Place}, \city{Hilo}, \state{HI} \postcode{96720}, \country{USA}}}

\affil[10]{\orgdiv{Department of Astronomy}, \orgname{University of Wisconsin---Madison}, \orgaddress{\street{475 N. Charter St.}, \city{Madison} \state{WI} \postcode{53706}, \country{USA}}}

\affil[11]{\orgdiv{School of Earth and Space Exploration}, \orgname{Arizona State University}, \orgaddress{\street{PO Box 876004}, \city{Tempe}, \state{AZ} \postcode{85287}, \country{USA}}}

\affil[12]{\orgdiv{Department of Physics}, \orgname{University of California, Santa Barbara}, \orgaddress{\street{} \city{Santa Barbara}, \state{CA} \postcode{93106}, \country{USA}}}

\affil[13]{\orgdiv{Department of Physics}, \orgname{University of California, Merced}, \orgaddress{\street{5200 North Lake Road}, \city{Merced}, \state{CA} \postcode{95343}, \country{USA}}}



\maketitle

\textbf{
In the contemporary Universe, most galaxies are supported by ordered rotation, yet a significant subset of the most massive and quiescent systems are dominated by random stellar motions and classified as slow rotators. 
These galaxies are widely thought to arise through processes that remove angular momentum and erase disk-like structures, but when and how this transformation occurs remains uncertain. 
Slow rotators are expected to be rare at early cosmic times, and observational studies of massive galaxies at high redshift have so far revealed only rapidly rotating systems. 
Here we report James Webb Space Telescope near-infrared integral field spectroscopy of XMM-VID1-2075, a massive quiescent galaxy at $z = 3.449$. 
The galaxy displays disturbed low-surface-brightness features and a low stellar spin parameter, $\lambda_{\rm R_e} = 0.123_{-0.023}^{+0.073}$, consistent with dispersion-dominated kinematics.
These results demonstrate that the formation of slow-rotating massive galaxies was already underway when the Universe was less than 2~Gyr old.}\\

Integral-field spectroscopy of nearby galaxies has established that, although rotational support dominates the galaxy population as a whole, a large fraction of the most massive systems are instead supported by stellar velocity dispersion and are classified as slow rotators \cite{Cappellari2007, Emsellem2011, Graham2018, Falcon-Barroso2019}. 
These galaxies are typically quiescent and are overrepresented in dense environments such as galaxy clusters, pointing to an evolutionary pathway distinct from that of disk-dominated systems \cite{Brough2017, Veale2018, Cole2020}. 
Measurements at low redshift further indicate that the abundance of slow rotators evolves only weakly with cosmic time for the last 4~Gyr, though may decrease before that, implying that the mechanisms responsible for their formation must act efficiently in the early stages of galaxy assembly \cite{Derkenne2024, Lopez2024}.

A leading explanation for the origin of slow rotators invokes repeated merger events that systematically lower stellar angular momentum and disrupt coherent rotational structures \cite{ Bois2011,Schulze2018}.
In this picture, the cumulative impact of mergers drives the transition from rotation-dominated progenitors to dispersion-supported remnants. 
Because massive galaxies at early epochs are expected to have experienced fewer mergers, theoretical models predict that pronounced slow rotators should be uncommon at high redshift, particularly beyond $z \gtrsim 3$  \cite{Khochfar2011, Kimmig2025}.

Direct observational tests of these predictions have been limited by the difficulty of obtaining spatially resolved stellar kinematics for massive quiescent galaxies at early times. 
Existing kinematic measurements of massive galaxies at high redshift consistently reveal systems with substantial rotational support \cite{Newman2018b, DEugenio2024, Pascalau2025}, and no galaxy has yet been confirmed as a slow rotator from stellar kinematics at $z \gtrsim 2$. 
This apparent absence has reinforced the view that the emergence of slow rotators is predominantly a late-time phenomenon.

In this work, we present near-infrared integral-field observations from the James Webb Space Telescope (\textit{JWST}) of XMM-VID1-2075, a massive quiescent galaxy at $z = 3.449$. We show that this galaxy exhibits both disturbed morphology indicative of past interactions and a low stellar spin parameter characteristic of slow rotators, providing direct evidence that merger-driven kinematic transformation was already occurring in the most massive galaxies at very early cosmic epochs.\\

XMM-VID1-2075 was initially selected from a catalog based on near-infrared observations from the VISTA Deep Extragalactic Observations (VIDEO) survey \cite{Jarvis2013} due to its apparent brightness ($m_K=20.80$ AB), high photometric redshift ($z_{\rm phot}>3$), and red spectral energy distribution.
Spectroscopic follow-up with the Keck/MOSFIRE instrument as part of the Massive Ancient Galaxies At $z>3$ NEar-Infrared (MAGAZ3NE) survey confirmed the galaxy's high redshift ($z_{\rm spec}=3.45$, when the Universe was only 1.8 Gyr old), large stellar mass \mbox{($M_* = 3.3^{+0.1}_{-0.3}\times10^{11}$} times the mass of the sun, M$_\odot$), and low star formation rate \mbox{($SFR<1$~M$_\odot$/yr)} \cite{Forrest2020b}.
Further analysis also found a large stellar velocity dispersion consistent with the large stellar mass measurement ($\sigma_{*,\rm R_e}=379^{+85}_{-53}$~km/s) \cite{Forrest2022}.
Such galaxies are important testbeds for theories of galaxy formation and evolution as they not only require a rapid buildup of stellar mass in the early cosmos via high SFRs, but also a rapid quenching mechanism hundreds of millions of years before the epoch at which they are observed \cite{Forrest2020a, Carnall2023, Glazebrook2024}.

\begin{figure}[b]
\centering
\includegraphics[width=\textwidth, trim=0in 6in 0in 0in]{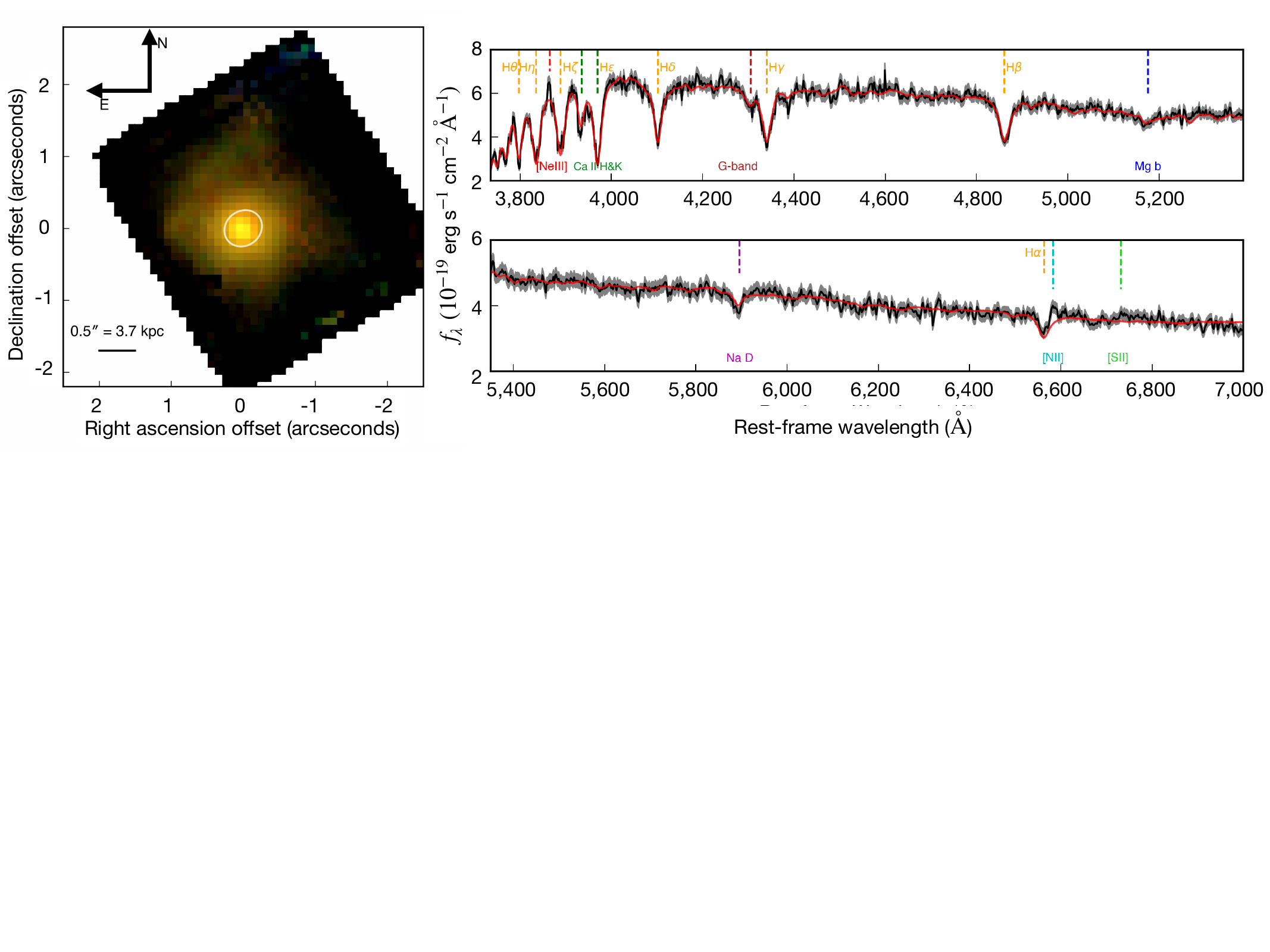}
\caption{\textbf{JWST/NIRSpec/IFU data of XMM-VID1-2075.} \textbf{Left:} An RGB image generated by collapsing the data cube along the wavelength axis (in three bins) to form a deep image. The half-light ellipse is shown in white. \textbf{Right:} The spectrum of all spaxels within the half light ellipse coadded together (black, with gray shading indicating uncertainties from adding in quadrature the fractional uncertainties of individual spaxels). The best-fit model from {\sc ppxf} is shown in red. Notable spectral features are labeled, including the Balmer series (hydrogen),
several Fraunhofer metal absorption features (\hbox{{\rm Ca}\kern 0.1em{\sc ii h\&k}},
G-band, \hbox{{\rm Mg}\kern 0.1em{b}}, \hbox{{\rm Na}\kern 0.1em{\sc d}}), 
and (very weak) emission lines (\hbox{{\rm [Ne}\kern 0.1em{\sc iii}{\rm ]}},
\hbox{{\rm [N}\kern 0.1em{\sc ii}{\rm ]}}, \hbox{{\rm [S}\kern 0.1em{\sc ii}{\rm ]}}).
There is no evidence of \hbox{{\rm [O}\kern 0.1em{\sc iii}{\rm ]}} emission at 5007~{\rm \AA}.}
\label{fig:IFU}
\end{figure}

\textit{JWST} NIRSpec Integral Field Unit (IFU) observations were taken as part of a Cycle 2 program (GO 2913, PI: Forrest) which targeted three of the most massive spectroscopically confirmed quenched galaxies at $z\sim3.5$, all similarly selected from the MAGAZ3NE survey (XMM-VID3-1120 and XMM-VID3-2457 with $M_*=3.0\times10^{11}$~M$_\odot$ and $M_*=1.8\times10^{11}$~M$_\odot$, respectively, both at $z_{\rm spec}=3.49$)\cite{Forrest2020b}.
\mbox{XMM-VID1-2075} was observed for 2.9h with the G235M/F170LP \mbox{grating/filter} combination which provides spectral coverage from $1.66 \leq \lambda/{\mu \rm m} \leq 3.17$ with a resolution of $R\sim1000$, corresponding to $\Delta v\sim300$~km/s.
The median signal-to-noise ratio (SNR) in the central spaxel for XMM-VID1-2075 is $SNR\sim40$ per wavelength element, or $SNR\sim17~{\rm \AA}^{-1}$ (rest-frame).

Galaxy kinematics from slit spectroscopy are typically measured within the half-light radius, $R_e$, within which half of the galaxy's luminosity is contained.
Spatially resolving this region for massive quiescent galaxies at $3<z<4$ where $R_e\sim 1-2$~kpc \cite{Straatman2015} requires angular resolutions of better than $\sim0.2^{\prime\prime}$.
The NIRSpec/IFU observations provide a spectrum in each $0.1'' \times 0.1''$ spaxel, equivalent to 730 pc on a side at this redshift.
While large ground-based observatories can reach this resolution with the aid of adaptive optics, the combination of excellent spatial resolution, spectral resolution, wavelength access, and sensitivity of the NIRSpec/IFU allows for kinematic modeling across spectral ranges of small galaxies which are impractical or impossible with any other facility.

In order to determine the half-light radius of XMM-VID1-2075, we collapse the IFU data cube along the spectral axis to create an image (left panel of Figure~\ref{fig:IFU}) noting that the system contains extended low surface brightness asymmetries, which in the local Universe are more commonly observed in massive galaxies and slow rotators \cite{Rutherford2024,Sola2025}, and are suggestive of merger activity.
We fit an ellipse which contains half the light of the galaxy in this collapsed image, which results in a semi-major radius \mbox{${\rm R_{e,maj}}=2.25\pm0.37$~kpc} and an ellipticity \mbox{$\epsilon=0.12\pm0.03$}, and thus a circularized effective radius of $R_{\rm e, circ} = 2.00$~kpc.

\begin{figure}[t]
\centering
\includegraphics[width=\textwidth, trim=0in 2in 0in 0in]{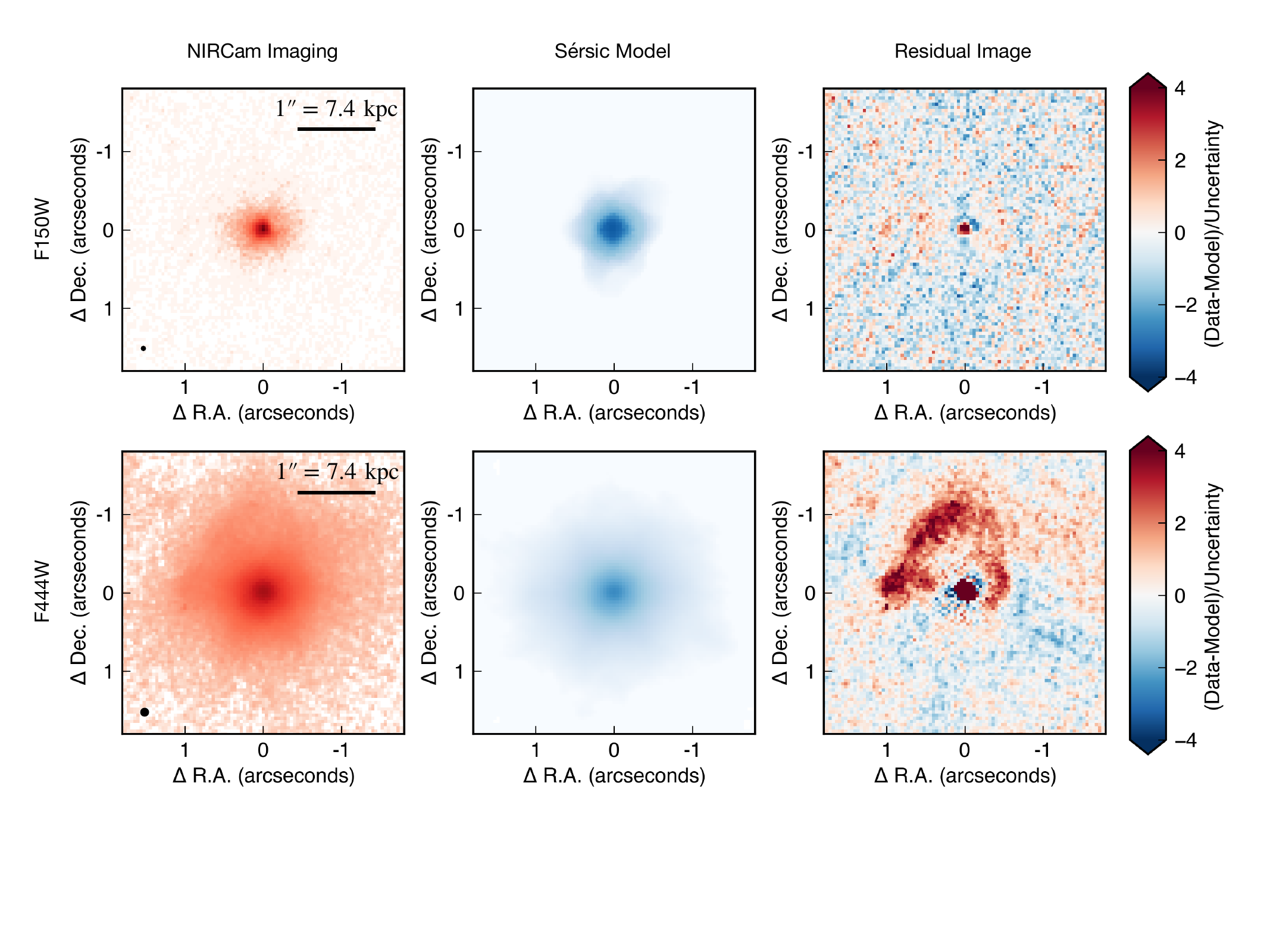}
\caption{\textbf{S\'ersic modeling of XMM-VID1-2075 from NIRCam imaging.} \textbf{Left:} NIRCam imaging flux, F150W on top, F444W on bottom, plotted on a log scale. The size of the PSF (FWHM) is shown as a black circle in the lower left. \textbf{Middle:} Single S\'ersic models fit to the observations, plotted on a log scale. \textbf{Right:} Residuals from the fits in units of image flux uncertainties, with excess from data in blue, plotted on a linear scale. Typical flux uncertainties per pixel are on the order of $\sim0.5$~nJy. There is clear evidence of additional flux in F444W to the NE of the center of the galaxy consistent with galaxy-galaxy interactions or merger activity.}
\label{fig:MORPH}
\end{figure}

As a check, we also perform S\'ersic modeling of NIRCam imaging of XMM-VID1-2075 which was acquired in a separate Cycle 2 program \cite{Ito2025}.
This results in a similar characterization of the half-light isophote, with $R_{\rm e} = 1.95 \pm 0.15$~kpc and $\epsilon=0.19\pm0.01$ depending on the bandpass.
Additionally, we find a S\'ersic index of $n=3.72 \pm 0.05$ and also note clear residuals in alignment with the low-surface brightness features seen in the collapsed IFU data cube consistent with merger activity (Figure~\ref{fig:MORPH}).
In the calculations and analysis below we use the half-light ellipse from the collapsed IFU data cube for consistency when comparing to the other two targets in the IFU observations, however the conclusions of the work do not change if the NIRCam fit is used instead.

\begin{figure}[b]
\centering
\includegraphics[width=\textwidth, trim=0in 2.5in 0in 0in]{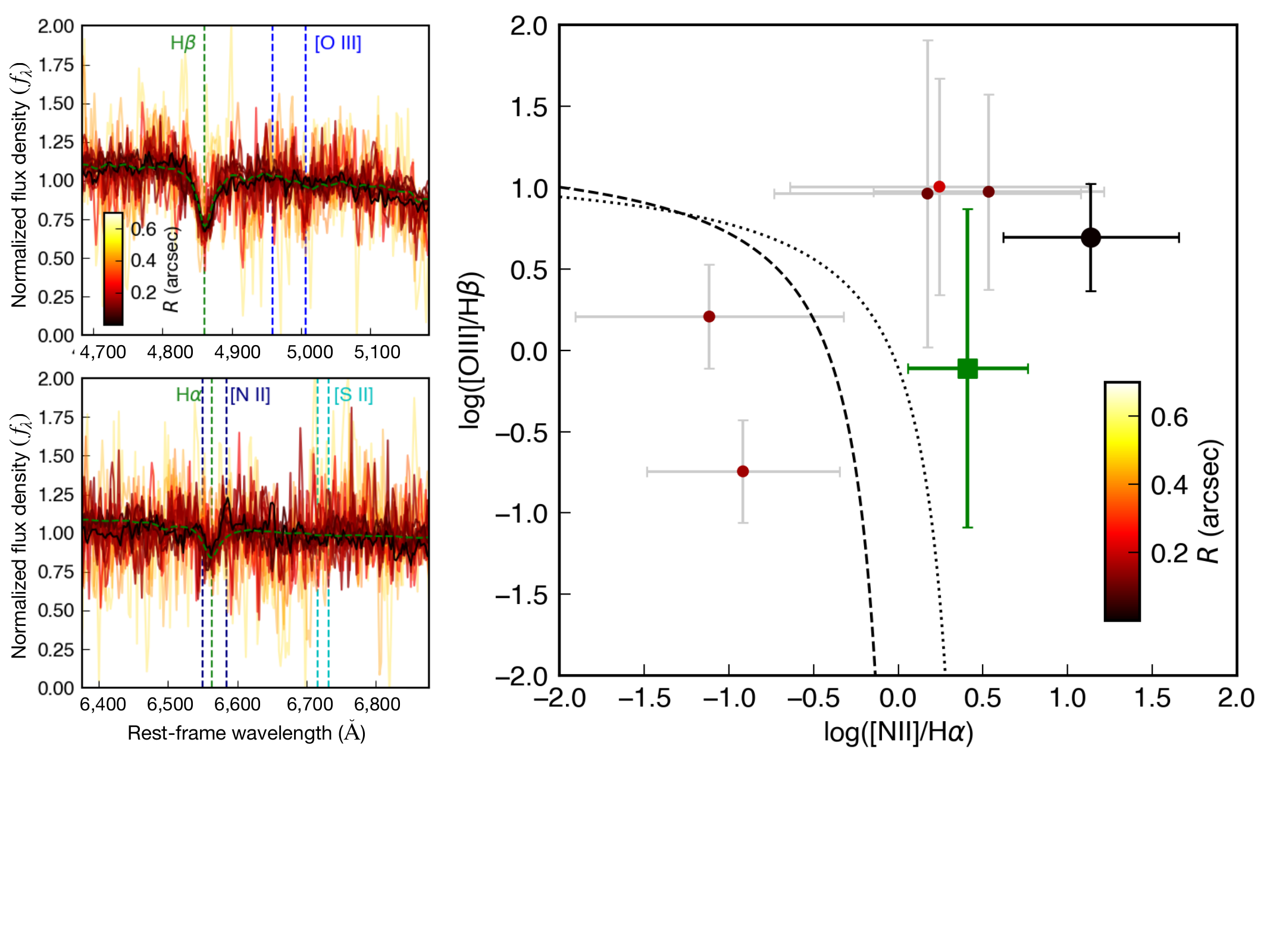}
\caption{\textbf{Analysis of emission origin via emission line ratios.} \textbf{Left:} The spectra around the H$\beta$ and \hbox{{\rm [O}\kern 0.1em{\sc iii}{\rm ]}} features (top) and the H$\alpha$ and \hbox{{\rm [N}\kern 0.1em{\sc ii}{\rm ]}} features (bottom) colored by the distance of the bin from the galaxy center. The best-fit stellar template to the spectrum of the entire galaxy is shown in green. 
\textbf{Right:} The Baldwin-Phillips-Terlevich diagram for discrimination between AGN and star-formation dotted line from \citet{Kewley2001}, dashed line from \citet{Kauffmann2003}). Only those Voronoi bins with line ratio uncertainties $<1$~dex (calculated by adding in quadrature the uncertainties on continuum level and line flux) are shown. Points are colored in the same manner as the left panel, and the green square represents the lines strengths from the fit to the spectrum of the entire galaxy. The central spaxel (black) shows evidence of \hbox{{\rm [N}\kern 0.1em{\sc ii}{\rm ]}} emission in the core of the galaxy consistent with an AGN.}
\label{fig:BPT}
\end{figure}

By coadding all the spaxels (including fractional weighting) within this ellipse, the effective spectrum within the half-light ellipse is generated (right panel of Figure~\ref{fig:IFU}).
This spectrum clearly shows absorption features from hydrogen and metals, and only very weak emission from ionized gas, in agreement with NIRSpec micro-shutter assembly spectroscopic observations \cite{Ito2025}.
The ratio of detected \hbox{{\rm [N}\kern 0.1em{\sc ii}{\rm ]} $\lambda 6584$} to H$\alpha$ $\lambda 6563$ in emission, as well as the detection of \hbox{{\rm [Ne}\kern 0.1em{\sc iii}{\rm ]} $\lambda 3869$}  emission is suggestive of an active galactic nucleus (AGN; Figure~\ref{fig:BPT}), although the complete lack of \hbox{{\rm [O}\kern 0.1em{\sc iii}{\rm ]}} emission, typically resulting from gas ionization by high-energy photons from the AGN, is unusual \cite{ Heckman1980b,Baldwin1981}, but not unheard of \cite{Agostino2023}.

The compact nature of XMM-VID1-2075 means that while the central spaxels of the galaxy have a sufficiently high SNR to accurately model stellar velocities and dispersions, spaxels at larger radii do not.
We therefore group spatially adjacent spaxels together using Voronoi binning 
so that each binned spectrum has a median $SNR\gtrsim20$ per wavelength element.
We compute the inverse-variance weighted sum of fluxes across the observed spectral range for each spaxel in a bin, and then weight the distance of each spaxel from the center of the galaxy to calculate a distance for the bin.

\begin{figure}[t]
\centering
\includegraphics[width=\textwidth, trim=0in 1in 0in 0in]{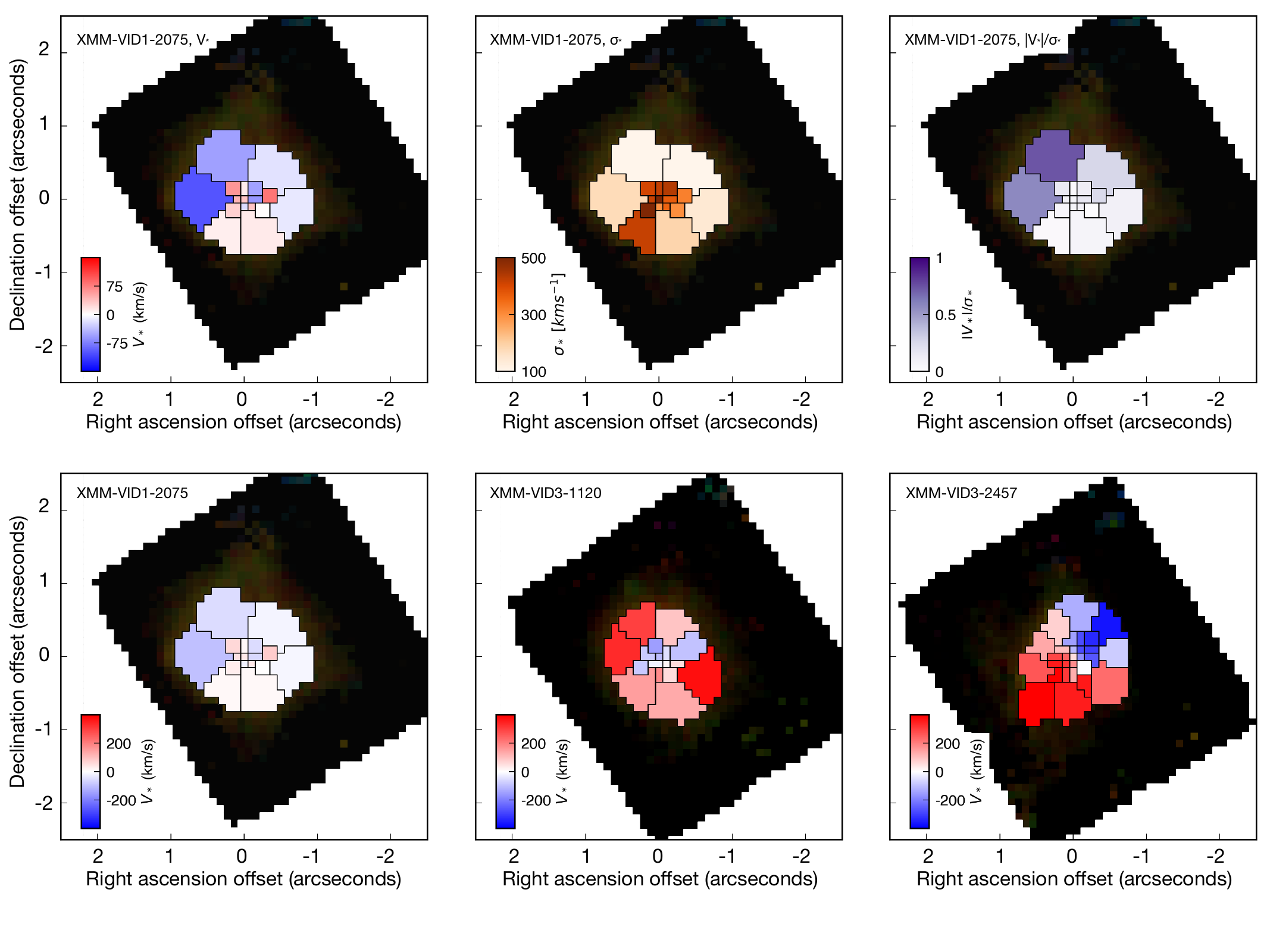}
\caption{
\textbf{Results of kinematic fitting to Voronoi binned data.}
\textbf{Top:} Kinematics of XMM-VID1-2075: best-fit stellar velocity offsets from the systemic velocity, $V_*$ (left),  best-fit stellar velocity dispersions, $\sigma_*$ (center), and the ratio of stellar velocity offset to velocity dispersion (right). All bins have \mbox{$|V_*|/\sigma_*<1$}.
\textbf{Bottom:} The best-fit stellar velocity offsets from systemic velocity for the 3 targets in GO-2913 using the same scaling in all panels (but different from the top left panel). From left to right: XMM-VID1-2075, XMM-VID3-1120, and XMM-VID3-2457.}
\label{fig:KIN_BIN}
\end{figure}

Spectral fitting is then performed on each bin \cite{Cappellari2023} 
to calculate a stellar velocity offset (\mbox{$V_* = v_*-v_{\rm sys}$}) and stellar velocity dispersion ($\sigma_*$) in each region (Figure~\ref{fig:KIN_BIN}), resulting in typical velocity uncertainties of 25 km/s.
The stellar velocity dispersion is approximately $\sigma_*\sim500$~km/s at the center of the galaxy, and that within ${\rm R_{e}}$ is measured to be $\sigma_{*,\rm R_{e}}=387\pm22$~km/s, consistent with the previous measurement from ground-based spectroscopy.
The stellar velocity offsets relative to the systemic velocity are remarkably small however, with all bins \mbox{$|V_*| < 100$~km/s} and a median absolute deviation \mbox{$MAD_V=26$~km/s}. 
Such low velocities, as well as the lack of a clear gradient to them suggest that this galaxy does not have significant ordered rotation.
This is in contrast with the other two massive, quiescent galaxies at $z\sim3.5$ with IFU observations taken in the same program.
Using the same analysis methods reveals that XMM-VID3-1120 and XMM-VID3-2457 have lower (but still large) central velocity dispersions of $\sim420$~km/s, and regions with increased rotational velocities up to $\sim 400$~km/s in the case of XMM-VID3-2457 (Figure~\ref{fig:KIN_BIN}).

\begin{table*}[t]
	\centering
	\caption{Galaxy properties derived from NIRSpec/IFU observations. Uncertainties take into account results from different model libraries ($z_{\rm spec}$) and different model PSFs ($r_{\rm SMA}$,  $\epsilon$, $\lambda_{r_e}$).}
	\label{tab:props}
	\begin{tabular}{cccccc}
		\hline
		Galaxy &		$z_{\rm spec}$     &   $r_{\rm SMA}$~(kpc)  &  $\epsilon$     & $\lambda_{r_e}$\\
		\hline
		\vspace{0.03in}
		XMM-VID1-2075 & $3.4490 \pm 0.0027$ 	& $2.25\pm0.37$ & $0.12^{+0.05}_{-0.03}$ & $0.123^{+0.073}_{-0.023}$ \\
		\vspace{0.03in}
		XMM-VID3-1120 & $3.4891 \pm 0.0026$ 	& $1.88\pm0.68$ & $0.18^{+0.02}_{-0.02}$ & $0.296^{+0.112}_{-0.071}$ \\
		\vspace{0.03in}
		XMM-VID3-2457 & $3.4892 \pm 0.0019$ 	& $2.08\pm0.31$ & $0.31^{+0.04}_{-0.03}$ & $0.671^{+0.048}_{-0.032}$ \\
		\hline
		\hline
	\end{tabular}
\end{table*}

The relative contributions of rotation and dispersion to the galaxy can be quantified via either a simple ratio of $|V_*|/\sigma_*$, as in the top right panel of Figure~\ref{fig:KIN_BIN}, or by calculating the spin parameter
\begin{equation}
\lambda_{\rm R_e} = \frac{\Sigma_{i=0}^N F_i r_i|V_i|}{\Sigma_{i=0}^N F_i r_i \sqrt{V_i^2+\sigma_i^2}}
\end{equation}
which considers the velocity ($V_i$) and dispersion ($\sigma_i$) in each bin weighted by the inverse-variance weighted flux summed across all observed spectral wavelengths ($F_i$) and distance from the center of the galaxy ($r_i$) for each of the $N$ bins with $r_i<R_e$.
Here we consider regions within the best-fitting half-light ellipse ${\rm R_{e}}$, but again note that differences in the choice of radius have no substantial effect on the calculated value.
We measure \mbox{$\lambda_{\rm R_e}=0.095$}, which is then corrected for the point spread function \cite{Harborne2020,DEugenio2024} and Voronoi binning bias to \mbox{$\lambda_{\rm R_e}=0.123^{+0.073}_{-0.023}$} (see Table~\ref{tab:props}).

\begin{figure}[t]
\centering
\includegraphics[width=0.75\textwidth, angle=270, trim=0in 0.5in 1.5in 0in]{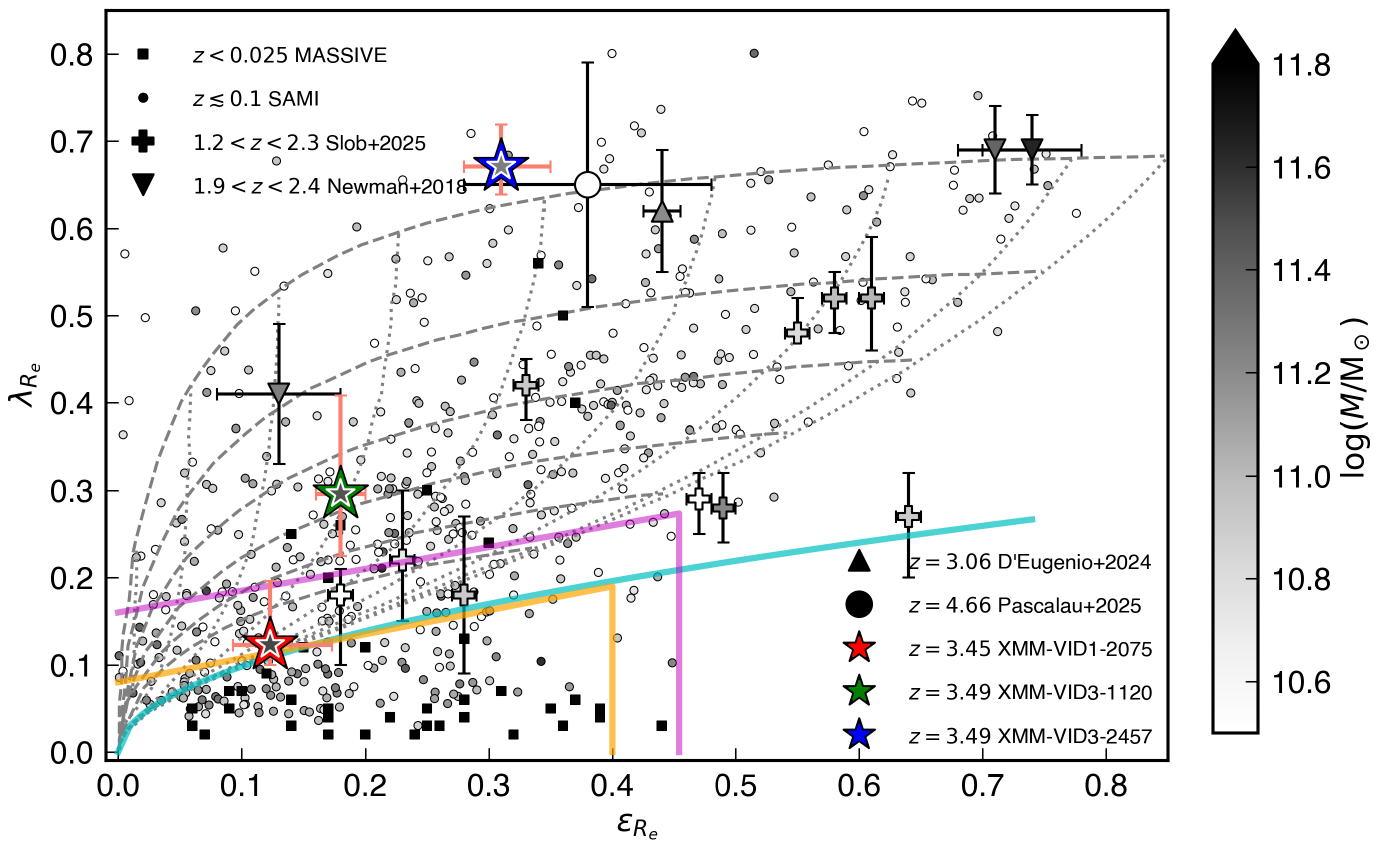}
\caption{\textbf{The spin parameter-ellipticity plane for massive ellipticals.} Low-redshift galaxies from the MASSIVE and SAMI surveys are shown as squares and small circles, respectively. High redshift observations with measured values of spin parameter are also included as crosses \cite{Slob2025}, downward-pointing triangles \cite{Newman2018b}, upward-pointing triangle \cite{DEugenio2024}, and large circle \cite{Pascalau2025}.
All points are shaded by their stellar mass. 
Error bars represent 1$\sigma$ uncertainties. 
The three targets in GO-2913 are stars outlined in color. The cyan, gold, and magenta curves show empirical cuts between fast- (above) and slow- (below) rotators \cite{Emsellem2011, Cappellari2016, vandeSande2021a}, while the gray lines show the observed changes for rotating galaxies with inclination varying from edge-on (right) to face-on (left) and intrinsic ellipticities (higher to lower moving from top to bottom). XMM-VID1-2075 stands out as the clearest high-redshift slow rotator.}
\label{fig:SPIN}
\end{figure}

In Figure~\ref{fig:SPIN} we show where massive galaxies over a range of redshifts lie on the $\lambda_{\rm R_e}-\epsilon$ plane.
The position of a galaxy on this plot is dependent not only on the amount of rotational support present in the galaxy, but also the intrinsic three-dimensional shape and the inclination of the galaxy relative to the observer's line of sight.
Depending on these parameters, fast-rotating galaxies generally lie within the grid of gray dashed and dotted lines, while slow rotators lie near the bottom of the plot.

XMM-VID1-2075 falls in a region which contains some slow rotators at low redshifts, however fast rotators which are aligned face-on to the observer can also lie in this region.
However, several lines of evidence suggest this is not the case for XMM-VID1-2075.
First, the morphological modeling of the NIRCam observations finds a S\'ersic index of  $n=3.72\pm0.05$ consistent with an elliptical galaxy.
Secondly, the observed axis ratio of the half-light ellipse can be used to derive an inclination assuming an intrinsic oblate spheroid structure consistent with a disk \cite{Hubble1926},
\begin{equation}
{\rm cos}^2 i = \frac{(b/a)^2-q_0^2}{1-q_0^2}
\end{equation}
where $i$ is the inclination angle of the galaxy axis relative to the observer, $b/a$ is the observed axis ratio, and $q_0$ is the axial ratio were the system to be observed edge-on.
Given the observed axis ratio of 0.89 ($=1-\epsilon$) and assuming $q_0=0.2$ appropriate for disk galaxies \cite{Turner2017}, 
the derived inclination angle is $i\sim28$ degrees, while assuming $q_0=0.7$, on the high end for ellipticals \cite{Mendez-Abreu2010} results in $i\sim40$ degrees.
At such an inclination, any significant rotation would be easily detected in our dataset.
Finally, the dynamical and stellar masses for XMM-VID1-2075 are both log($M$/M$_\odot$)$\sim11.5$, whether calculated from \textit{HST} imaging and Keck spectroscopy \cite{Forrest2022} or from \textit{JWST} observations, while in the case of a face-on rotator, using the line-of-sight velocity dispersion to calculate dynamical mass would lead to an underprediction \cite{Bellovary2014}.
We therefore conclude that XMM-VID1-2075 is not a face-on fast-rotator.

Of the galaxies at $z>1.2$ plotted in Figure~\ref{fig:SPIN}, four have reported stellar masses log($M$/M$_\odot$)~$\gtrsim11.4$:
1) MRG-M0138 \cite{Newman2018b} at z=1.95 with clear rotation ($\lambda_{\rm R_e}=0.69$);
2) XMM-VID3-1120 \cite[][this work]{Forrest2020a} at z=3.49 whose kinematics suggest a potential merger instead of disk rotation (measured $\lambda_{\rm R_e}=0.25$, Figure~\ref{fig:KIN_BIN});
3) XMM-VID1-2075; and
4) JWST-SUSPENSE 130647 \cite{Slob2025} at z=1.51, though this galaxy only has a lower limit on rotation due to potential alignment issue with the slit observations ($\lambda_{\rm R_e}>0.08$).
Thus, while the majority of the massive galaxies, log($M$/M$_\odot$)~$\gtrsim10.5$, at $z>1.2$ with stellar kinematic measurements are rotating, potentially 3/4 of the most massive galaxies, log($M$/M$_\odot$)~$\gtrsim11.4$, do not have significant rotation, suggesting that even at early epochs, slow rotators may be prevalent among this extreme population.

As can be seen in Figure~\ref{fig:SPIN}, other high-redshift massive galaxies at $z>3$ with IFU observations \cite{DEugenio2024, Pascalau2025}, including the other two galaxies in this program, have large measured rotational velocities (and therefore large values of $\lambda_{\rm R_e}$), indicating that any merger activity which may have taken place in their formation has not yet had a significant effect on their kinematic evolution.
Massive galaxies at $z\gtrsim2$ with rotational velocities measured from slit spectroscopy \cite{Newman2018b, Slob2025} also have large $\lambda_{\rm R_e}$, though determinations of rotation from such data are more susceptible to projection and alignment effects and cannot confirm a lack of rotation.
In contrast to all these galaxies, XMM-VID1-2075 is dispersion-dominated, with evidence for only minimal rotation.

In fact, XMM-VID1-2075 is more similar kinematically to the most massive early-type galaxies in the local Universe than other observed high-redshift galaxies, though it is smaller in size than the local slow rotators.
In the canonical picture of formation for a low-redshift `dead' elliptical, a compact, massive progenitor undergoes numerous minor mergers (mass ratios $\gtrsim4:1$)
which cumulatively remove angular momentum from the system \cite{VanDokkum2010}.
This process also can increase the size and S\'ersic index of the system and yield a rounder galaxy than the elongated slow rotators expected in a major merger scenario \cite{Naab2014}.
In this way, numerous minor mergers could form something similar in stellar mass and kinematics to XMM-VID1-2075, though as mentioned above, this would be expected to produce a galaxy larger in size.
Indeed, while the low-surface brightness feature on the NE side of XMM-VID1-2075 could be indicative of a recent such minor merger, it is unlikely that sufficient minor merging events to build the observed stellar mass could occur in the first 1.8 Gyr of cosmic time, particularly without leaving additional morphological signatures.
We thus explore alternative formation scenarios.

One such alternative picture of formation for massive, quenched ellipticals includes a major merger (roughly equivalent progenitor masses) as the defining event.
At \mbox{$\sim500$~Myr} post-merger, 1) any signatures of black hole growth (active galactic nucleus, AGN) have rapidly faded, 2) new star-formation has ceased, and 3) tidal features from the merger have also faded and are only visible with very deep observations, aside from which the galaxy is very compact \cite{Hopkins2008}.
XMM-VID1-2075 matches these characteristics.
In order for such a system to show a lack of rotation however, the two progenitors must have anti-aligned angular-momentum vectors of similar amplitude which is not expected to be common.
It is unclear how prevalent galaxies similar to XMM-VID1-2075 are at this epoch, and larger samples are therefore necessary to judge if such an explanation is valid.
It is also unclear if the low-surface brightness feature on the NE side of XMM-VID1-2075 is indeed the result of a major merger, as the flux at the center of the galaxy is large relative to that contained in this structure ($>10:1$).
That said, studies of local massive early-type galaxies have revealed that their cores consist of older stellar populations with larger velocity dispersions than the outlying regions \cite{VanDokkum2017, LaBarbera2019}.
If XMM-VID1-2075 undergoes additional minor merger activity, which tends to deposit stars in the outskirts of the system \cite{Karademir2019}, its size would increase significantly.
Combined with only small increases expected in stellar mass and velocity dispersion, this would result in a descendant galaxy akin to local massive slow rotators \cite{Bezanson2009, Saracco2020, Forrest2022}, though this may also require accretion of `mini-mergers' into the galaxy envelope \cite{Suess2023, Nipoti2025}.

A third potential explanation is that the apparent merger and the kinematic state of XMM-VID1-2075 are independent.
Simulations have suggested that a massive galaxy at early times can have rotation removed via an isotropic infall of gas.
In this scenario, the kinematic transformation from fast- to slow-rotator occurs before/during the starburst phase, which in turn triggers AGN and supernova feedback to quench the galaxy - 3/35 quiescent galaxies from the Magneticum simulation at $z=3.42$ are slow rotators, all of which were quenched in this way \cite{Kimmig2025}.
A subsequent dry (gas-poor) minor merger which produces the low-surface brightness feature could then be the first of many which will occur throughout the continued evolution of the system toward a massive `dead' elliptical system similar to those seen in the local Universe.
Discriminating between these potential formation pathways for XMM-VID1-2075 is difficult and requires deeper data to analyze additional signatures such as higher-order kinematic moments (skewness, kurtosis) \cite{Naab2014, Forbes2017, VandeSande2017}.
Larger samples of massive quiescent galaxies with stellar rotation measurements will also enable statistical discrimination between the anti-aligned merger and gas inflow kinematic transformation pathways.

XMM-VID1-2075 is the first slow rotator confirmed via stellar kinematics at $z\gtrsim2$.
Simulations have offered different predictions on whether slow rotators can \cite{Kimmig2025} or can not \cite{Penoyre2017} form at this early epoch, as well as the mechanisms behind their eventual formation, but all agree that the vast majority of slow-rotators form at $z\lesssim1-2$ \cite{Lagos2017}.
Discerning between possible formation pathways such as isotropic gas inflow (potentially common) and a counter-rotating major merger (likely rare) requires resolved kinematic studies of additional massive, quiescent galaxies in the first few Gyr of cosmic time.
Given the capabilities of the JWST/NIRSpec/IFU and the increasing numbers of spectroscopically confirmed massive, quiescent galaxies in the early Universe, building a large sample of similar galaxies with spatially-resolved spectroscopy is now possible.
Such a sample will allow for analyses of rotational support in conjunction with the prevalence of morphological disturbances, allowing for statistical insights into the role that different mechanisms such as mergers and AGN activity play in the formation, quenching, and kinematic and morphological transformations of massive galaxies in the early Universe.

\clearpage

\section{Methods}\label{sec11}

\subsection{JWST Observations}

\subsubsection{NIRSpec/IFU Data}

JWST/NIRSpec/IFU \cite{Boker2022} data were taken on August 2, 2024 as part of a program targeting ultra-massive quiescent galaxies at $z\sim3.5$ (GO-2913; PI: B. Forrest).
The JWST/NIRSpec/IFU is an integral field unit spectrograph with a $3'' \times 3''$ field of view and spaxels $0.1''$ on a side and is the first IFU capable of resolving these compact objects.
Observations used the G235M/F170LP grating and filter combination and a four-point dither pattern and NRSIRS2 readout pattern, with a total integration time of 2.9 hours.
Initial data reduction was performed with the \textit{JWST} pipeline version 1.14.0 and context file \textit{jwst\_1256.pmap}.
This pipeline consists of three stages, which perform many tasks including correcting for detector sensitivity variations, performing ramp-fitting, flat-fielding, flagging pixels affected by MSA shutters which are stuck open, outlier detection, 3D cube-building, and 1D spectral extraction.
Additional code was written to mask contaminating flux, more strictly flag cosmic rays and hot pixels, and perform background subtraction.
In each exposure, we identify pixels which satisfy all of the following: 1) a flux greater than $2\sigma_{\rm NMAD, im}$ above the median of the entire image, 2) a flux greater than $2\sigma_{\rm NMAD, spec}$ above the median flux of the spectrum in that spaxel, and 3) an adjacent pixel flagged by the pipeline.
We confirm that what minimal line emission exists in this target is not affected by this process.
To determine background signal, we calculate the median spectrum in twenty off-source spaxels and then smooth with a Savitzky-Golay filter with window length 20 and polynomial order 3 to generate a background spectrum which is subtracted from all spaxels.
These corrections are necessary to reveal low surface brightness features.

\subsubsection{NIRCam Data}

JWST/NIRCam observations were taken on December 30, 2023 (0.4h in F150, F200, F356W, and F444W) as part of the DeepDive program observing eight massive quiescent galaxies at $3.5 \lesssim z \lesssim 4.0$ \cite[GO-3567, PI: F. Valentino; ][]{Ito2025}.
Data were taken in the F150W, F200W, F356W, and F444W bandpasses, with integration times of $\sim0.4$h per band, with a pixel scale of $0.03-0.06"$/pixel depending on the filter.
Reduced data were retrieved from the Dawn JWST archive, which processed the images with \textsc{grizli} \cite{grizli_1_9_11, Valentino2023}.
For XMM-VID1-2075, these observations achieve $SNR\sim80$ for the F150W filter which is blueward of the Balmer break and $SNR\gtrsim200$ for the other three filters.

\subsection{Redshift Determination}

We employ a Monte Carlo methodology to determine the systemic redshift of the galaxy, which is critical for accurately measuring kinematic properties.
For each iteration, each spectral element in each spaxel is drawn from a normal distribution centered at the observed flux with standard deviation equivalent to the uncertainty on the observed flux.
We then coadd the resulting spectra of all spaxels with median signal-to-noise ratio $\tilde{SNR}>10$ to generate a `total' spectrum of the galaxy.
This spectrum is then fit using the penalized pixel-fitting ({\sc ppxf}) method \cite{Cappellari2017, Cappellari2023}, which accounts for the resolution of the instrument while modeling the stellar continuum as a superposition of simple stellar population spectra.
We use four different model libraries: E-MILES \cite{Vazdekis2016}, FSPS \cite{Conroy2009, Conroy2010}, GALAXEV \cite{Bruzual2003}, and XSL \cite{Verro2022}, minimizing the velocity offset of the galaxy spectrum for each set.
The median redshift from all libraries and iterations is \mbox{$\tilde z = 3.44898 \pm 0.00057$ (statistical) $^{+0.00039}_{-0.00258}$ (systematic)}.
We also confirm that the resulting fit velocity dispersion is the minimum value at this redshift for each model set.
However, given the sensitivity of kinematic results to the systemic redshift, in this analysis we use results from the EMILES libraries for consistency, with a best fit redshift of $z_E=3.44927$.

As a check, the total spectrum was also fit in conjunction with photometry from catalogs based on near-infrared observations from the VIDEO survey \cite{Jarvis2013} using FAST++ \cite{Schreiber2018a} with the BC03 models \cite{Bruzual2003}.
This results in a best-fit redshift of $z_{\rm F++}=3.4461$.
We note that all redshifts are consistent with the previous determinations from Keck/MOSFIRE ($z=3.4523$) and Keck/NIRES ($z=3.4482$) spectra \cite{Forrest2022}, though in better agreement with the latter.
Additionally, analysis of a JWST/NIRSpec/MSA spectrum of XMM-VID1-2057 resulted in a redshift $z=3.4472$. \cite{Ito2025}

Finally, we compare the wavelengths of Balmer absorption lines in the central spaxel to the best-fit redshift model and find a scatter of 8~km/s around the expected rest-frame wavelengths, indicating that the wavelength calibration uncertainties in these data are considerably smaller than other factors in redshift determination.
The results of this work are insensitive to different choices of redshift over this narrow range.

\subsection{Spatial Binning and Kinematic Modeling}

The {\sc ppxf} method used for determining the redshift is also used for modeling the stellar line-of-sight velocity distribution across the projected face of the galaxy.
Spaxels are spatially binned using a Voronoi binning technique \cite{Cappellari2003} to obtain $SNR\gtrsim20$ per bin (bins are shown in Figure~\ref{fig:KIN_BIN}).
This was chosen as a balance between having sufficient bins to map the face of the galaxy and sufficient $SNR$ to reliably measure kinematics.

The center of each galaxy was taken to be the average inverse-variance flux weighted position in the collapsed data cube:
\begin{eqnarray}
\bar{x} = \frac{\Sigma_{i=0}^{N} x_i(F_i/\delta F_i^2)}{\Sigma_{i=0}^N F_i/\delta F_i^2},
\end{eqnarray}
where \textit{x} is either right ascension or declination, \textit{N} is the number of spaxels in the dimension in question, $F_i$ is the total flux in a spaxel collapsed over the entire spectral wavelength, and $\delta F_i$ is the uncertainty on $F_i$.

The calculation of the distance of each binned group of spaxels from the center was calculated in the same flux-weighted manner, where \textit{N} is the number of spaxels in a given bin and \textit{x} is the distance of that spaxel from the galaxy center.

The $\lambda_{\rm R_e}$
measurement is insensitive to the choice of Voronoi binning of spaxels provided, but the uncertainties on both velocity offset and stellar velocity dispersion increase significantly when binning is done with a target $SNR\lesssim15$.
We note that the same choice of $SNR$ does reveal clear kinematic patterns in the other two targets in this program (Figure~\ref{fig:KIN_BIN}).

\subsection{Modeling Galaxy Morphology}

We use \texttt{pysersic} \cite{Pasha2023} to model the light profile of XMM-VID1-2075 in all four filters of available NIRCam imaging (F150W, F200W, F356W, F444W).
The NIRCam model PSF was constructed from several nearby unsaturated and uncontaminated stars. For each star, nearby objects in an image cutout were masked, sky was subtracted, and integrated flux was normalized, before the resulting cutouts were stacked.
Three galaxy models were tested, including a single S\'ersic profile, a two-component S\'ersic profile, and a model with a S\'ersic profile and an additional point source.
Neither of the latter two options statistically improve the fit relative to the single S\'ersic model, which we refer to here.
The fits in all filters result in a S\'ersic index $3.67 \leq n \leq 4.07$, all with uncertainties $\delta n<0.15$.
Measured effective radii range from $1.72 \leq r_{\rm eff}/{\rm kpc} \leq 2.12$, compared to 1.81~kpc measured from the collapsed IFU data cube.
The ellipticity is measured as $0.18 \leq \epsilon \leq 0.21$ with uncertainty $\delta \epsilon \sim 0.02$, compared to $\epsilon=0.12\pm0.03$ from the IFU data.
The differences in effective radius and ellipticity between the NIRCam imaging and collapsed NIRSpec IFU data cube are small, but can be attributed to the poorer spatial sampling of the IFU which blurs the image quality slightly.
Additionally, the NIRSpec/IFU PSF is less well constrained from available data than that of the NIRCam instrument. We use several model PSFs, including those presented in \citet{DEugenio2024} and \citet{Bentz2025} while performing morphological fits and do not find significant differences between these models affecting the results of this work.
We note that using the NIRCam morphological parameters instead of those from the IFU data does not change the conclusions of this work.
Additionally, the model-subtracted imaging highlights the extended low-surface brightness asymmetries and is further suggestive of merger activity (Figure~\ref{fig:MORPH}).

\subsection{Emission and Absorption Line Fitting}

We attempt to fit any emission from H$\beta$, \hbox{{\rm [O}\kern 0.1em{\sc iii}{\rm ]}}, H$\alpha$, \hbox{{\rm [N}\kern 0.1em{\sc ii}{\rm ]}} and \hbox{{\rm [S}\kern 0.1em{\sc ii}{\rm ]}} in each binned spectrum, as well as the total galaxy spectrum.
The \hbox{{\rm [O}\kern 0.1em{\sc ii}{\rm ]}} doublet falls just blueward of the observed spectrum and is therefore not fit.
For each spectrum, as well as each spectrum with the best-fit stellar continuum model subtracted, we fit H$\beta$ and \hbox{{\rm [O}\kern 0.1em{\sc iii}{\rm ]}} emission with a triple Gaussian model, with each component forced to have the same velocity offset and velocity width and the line ratio of the \hbox{{\rm [O}\kern 0.1em{\sc iii}{\rm ]}} doublet fixed to 1:3.
We also fit H$\alpha$ and \hbox{{\rm [N}\kern 0.1em{\sc ii}{\rm ]}} with a similar triple Gaussian model, with the \hbox{{\rm [N}\kern 0.1em{\sc ii}{\rm ]}} line ratio fixed to 3:10.
\hbox{{\rm [S}\kern 0.1em{\sc ii}{\rm ]}} is fit as a double Gaussian as well.
While these features can have differing line widths and velocity offsets, particularly when an AGN is present, the weak emission lines seen here do not require inclusion of these additional degrees of freedom for a sufficiently good fit, and in fact there is no evidence for line emission for some or all of the fitted lines in most spaxels (Figure~\ref{fig:BPT}).

In the central spaxels, there is clear \hbox{{\rm [N}\kern 0.1em{\sc ii}{\rm ]}} emission as well as line infilling by H$\alpha$ emission, and the resultant line ratios are large enough to suggest the presence of a weak AGN.
However even at the center, there is no \hbox{{\rm [O}\kern 0.1em{\sc iii}{\rm ]}} or H$\beta$ emission of note.
\hbox{{\rm [N}\kern 0.1em{\sc ii}{\rm ]}} emission is also common in low ionization nuclear emission-line regions (LINERs), however other associated features including \hbox{{\rm [O}\kern 0.1em{\sc i}{\rm ]}} are not detected significantly in the data.
While recent work has also suggested that strong line ratios are not as effective at distinguishing AGN from star-formation at high redshifts due to degeneracies
between low metallicity stellar populations and accreting black holes \cite{Cleri2025}, as a massive, quiescent galaxy XMM-VID1-2075 does not contain such low-metallicity populations, and has line ratios consistent with other observed high-redshift massive quiescent galaxies \cite{Belli2017b, Kriek2024a, Bugiani2025}.
Furthermore, the two bins which lie in the nominal star-forming region of the Baldwin-Phillips-Terlevich diagram have exceedingly small (if any) emission, and thus any ongoing star formation is minimal.
We rule out significant dust attenuation obscuring an AGN (but not most stars), as neither very deep ALMA Band 7 observations \citep[0.87~mm / 344~GHz, rms = 11.8 ${\rm \mu}$Jy/beam;][]{Chang2026b} nor deep radio observations from MeerKat (1.2~GHz, rms = 5.1 ${\rm \mu}$Jy/beam) as part of the MIGHTEE survey \cite{Jarvis2016,Hale2025} detect anything at the position of XMM-VID1-2075.

\clearpage

\backmatter

\bmhead{Data availability}
	The JWST/NIRSpec/IFU observational data that support the findings of this study are publicly available from \href{https://archive.stsci.edu/}{https://archive.stsci.edu/} and can be found by searching for Program number=2913 and PI Surname=Forrest.
	
\bmhead{Code availability}
	The programming language Python \cite[\href{www.python.org}{www.python.org}][]{VanRossum1995} was used to analyze the data in this work. Specific packages used include
	Astropy \cite[\href{https://astropy.org}{https://astropy.org}][]{Astropy2022},
	Matplotlib \cite[\href{https://matplotlib.org}{https://matplotlib.org}][]{Matplotlib2007},
	NumPy \cite[\href{https://numpy.org}{https://numpy.org}][]{Numpy2020},
	and SciPy \cite[\href{https://scipy.org}{https://scipy.org}][]{Scipy2020}.

\bmhead{Acknowledgements}
This work is based on observations made with the NASA/ESA/CSA James Webb Space Telescope.
The data were obtained from the Mikulski Archive for Space Telescopes at the Space Telescope Science Institute, which is operated by the Association of Universities for Research in Astronomy, Inc., under NASA contract NAS 5-03127 for JWST.
These observations are associated with program \#GO-02913.
Support for program \#GO-02913 was provided by NASA through a grant from the Space Telescope Science Institute, which is operated by the Association of Universities for Research in Astronomy, Inc., under NASA contract NAS 5-03127.
BF acknowledges support from JWST-GO-02913.001-A.
GW gratefully acknowledges support from the National Science Foundation through grant AST-2347348. 
Some of the data products presented herein were retrieved from the Dawn JWST Archive (DJA). DJA is an initiative of the Cosmic Dawn Center (DAWN), which is funded by the Danish National Research Foundation under grant DNRF140.
Supported by the international Gemini Observatory, a program of NSF NOIRLab, which is managed by the Association of Universities for Research in Astronomy (AURA) under a cooperative agreement with the U.S. National Science Foundation, on behalf of the Gemini partnership of Argentina, Brazil, Canada, Chile, the Republic of Korea, and the United States of America.

\bmhead{Author contributions}
	B.F., A.M., D.M., M.C.C., and G.W. wrote the JWST proposal.
	B.F. reduced the data, with help from A.H.E. and J.A.-D.
	B.F. and A.M. led the interpretation.
	D.M., R.P., and N.O. performed the morphological fitting.
	B.F. wrote the manuscript text.
	All authors (B.F., A.M., D.M., R.P., N.O., J.A.-D., W.C., M.C.C., A.H.E., P.G., L.K., B.C.L., I.M., A.N., R.-S.R., S.M.U.S., G.W., and M.E.W.) contributed to analysis and interpretation of the data.

\bmhead{Competing interests}
	The authors declare no competing interests.

\clearpage




\end{document}